\documentclass[pdftex,twocolumn,epjc3]{svjour3}          

\RequirePackage[T1]{fontenc}

\smartqed  

\RequirePackage{graphicx}
\RequirePackage{mathptmx}      
\RequirePackage{flushend}
\RequirePackage[numbers,sort&compress]{natbib}
\RequirePackage[colorlinks,citecolor=blue,urlcolor=blue,linkcolor=blue]{hyperref}

\journalname{Eur. Phys. J. C}

\begin{document}

\title{Topological Lifshitz phase transition in effective model of QCD with chiral symmetry non-restoration}


\author{Tran Huu Phat\thanksref{e1,addr1,addr2}
        \and
        Phung Thi Thu Ha\thanksref{addr3,addr1}
		\and
        Nguyen Tuan Anh\thanksref{e2,addr4} 
}

\thankstext{e1}{e-mail: tranhuuphat41@gmail.com}
\thankstext{e2}{e-mail: dr.tanh@gmail.com}

\institute{Vietnam Atomic Energy Institute, 59 Ly Thuong Kiet, Hanoi, Vietnam\label{addr1}
          \and
          Center for Advance Study, Dong Do University, 170 Pham Van Dong, Hanoi, Vietnam 			\label{addr2}
          \and
          Institute of Physics, 10 Dao Tan, Hanoi, Vietnam\label{addr3}
	    \and
          Electric Power University, 235 Hoang Quoc Viet, Hanoi, Vietnam\label{addr4}
}

\date{Received: date / Accepted: date}

\maketitle

\begin{abstract}
The topological Lifshitz phase transition is studied systematically within an effective model of QCD,  in which the chiral symmetry, broken at zero temperature, is not restored at high temperature and/or baryon chemical potential. It is found that during phase transition the quark system undergoes a first-order transition from  low  density fully-gapped state to high density state with Fermi sphere which is protected by momentum-space topology. The Lifshitz phase diagram in the plane of temperature and baryon chemical potential is established. The critical behaviors of  various equations of state are  determined.
\end{abstract}

\section{Introduction}

Quantum Chromodynamics (QCD) is actually  considered  to be the theory of strongly interacting systems of quarks and gluons. At finite baryon chemical potential the theory also provides information on nuclear properties. We have  expected that  QCD  at finite temperature $T$ and baryon chemical potential $\mu$  has a very rich phase structure \cite{r1} such as  the chiral symmetry restoration at high temperature and/or  baryon chemical potential, the thermal phase transition from the Nambu-Goldstone phase to quark - gluon plasma, the phase transition from the Nambu-Goldstone phase to the color superconductivity at low  $T$  and so on. The experimental exploration of these phase transitions is being actively  carried out in RHIC (Relativistic Heavy Ion Collider) and then in LHC (Large Hadron Collider). Especially, the quantum phase transition from Nambu-Goldstone phase to the color superconductivity at low $T$, which is of interest in the interiors of neutron stars and possible quark stars,  is also relevant for experimental realizations with moderate energy  heavy-ion collisions. In addition to these phase transition patterns, based on the fundamental works of Weinberg \cite{r2}, Dolan and Jackiw \cite{r3} one believes that  generally symmetry will get restored at high temperature from the broken phase at $T = 0 $. However, in reality there exist physical systems which manifest the symmetry non-restoration (SNR) at high temperature. They associate with numerous different materials \cite{r4}. The theoretical  calculations performed in  Refs.~\cite{r5,r6}  proved that SNR  really occurs in several models. The physical significance of these phenomena rests in the fact that SNR may have remarkable consequences for cosmology, namely, within the scenario of  SNR  the puzzle relating to topological defects in the Standard Cosmological Big Bang of universe could be solved \cite{r5,r6}. Concerning   QCD the results derived from an effective $U(N_f)\times U(N_c)$ chirally invariant Nambu--Jona-Lasinio model coupled to a uniform temperature dependent $A_0$  gauge field \cite{r7} showed  that the chiral symmetry is not restored at the deconfining temperature in the case when  Re[tr$_c(P)$]$<0$, here $P$ is the Polyakov loop; this agrees with the simulation study for a quenched finite temperature QCD \cite{r8}. Until now the phase structure of QCD has been step by step established by mean of  either Lattice QCD \cite{r9} or  effective models of QCD for the scenario of  chiral symmetry restoration at high temperature and/or baryon density. Nevertheless, there remains lacking information  on  phase structure of QCD    corresponding to  non-restoration scenario.

In view of what presented above,  the present paper aims at remedying  this gap, namely, we  focus on investigating quantum phase transition of QCD  associating with the sector of  chiral symmetry non-restoration.  Quantum phase transition means the transition occurs due to  fluctuations driven by  the Heisenberg uncertainty  principle even in the ground state. For this purpose, we start from an effective model, whose  Lagrangian reads
\begin{eqnarray}
{\cal L} &=& \bar{\psi} [\gamma^\mu(i\partial_\mu \!-\! g_v\omega_\mu) \!-\! m_q \!+\! g_s(\phi \!+\! i\gamma^5\vec{\tau}.\vec{\pi})]\psi \nonumber\\
& &+ \frac{1}2 (\partial_\mu\phi\partial^\mu\phi \!+\! \partial_\mu\vec{\pi}\partial^\mu\vec{\pi}) -\frac{1}4 F_{\mu\nu}F^{\mu\nu}  -U,
\label{1}
\end{eqnarray}
where
\begin{eqnarray*}
U \!&=&\! -\frac{1}2 g_{sv}(\phi^2 \!+\! \vec{\pi}^2)\omega_\mu\omega^\mu + \frac{m^2}2 (\phi^2 \!+\! \vec{\pi}^2) + \frac{\lambda}{4!}(\phi^2 \!+\! \vec{\pi}^2)^2,\\
\lambda \!&=& \frac{3}{f_\pi^2}(m_\sigma^2 \!-\! m_\pi^2), \\
m^2 \!&=&  \frac{1}2 (3m_\pi^2 \!-\! m_\sigma^2).
\end{eqnarray*}
In ({\ref{1}) $F_{\mu\nu}=\partial_\mu\omega_\nu-\partial_\nu\omega_\mu$; $\psi$, $\phi$, $\vec{\pi}$, $\omega_\mu$ are respectively the field operators of  quark, sigma meson, pion and omega meson; $m_q$, $m_\sigma$, and $m_\pi$ are respectively masses of  current quark, sigma meson, pion and omega meson; $m$, $\lambda$, $g_s$, $g_v$, $g_{sv}$ are the model parameters.

The paper is organized as follows. The Section 2 is devoted to the calculation of effective potential and its related physical quantities. The chiral symmetry non-restoration is dealt with  in Section 3. In Section 4 we focus on the study of Lifshitz phase transition. The conclusion and discussion are  presented in Section 5.

\section{Effective potential and its related physical quantities}

Starting from the Lagrangian (\ref{1}) let us  calculate the effective potential and its related quantities.

In the mean field approximation 
\[
\langle \phi \rangle = u - \sigma_0, \;\;\;\;\;\;\; \langle \pi_i \rangle = 0 \;\;\;\;\;\;\; \langle \omega_\mu \rangle = \omega_0 \delta_{0\mu},
\]
the Lagrangian (\ref{1}) takes the form
\begin{eqnarray}
{\cal L} =  \bar{\psi} [\gamma^0 (p_0 + \mu^*) + \vec{\gamma}.\vec{p} -m_q^*]\psi  -U,
\label{2}
\end{eqnarray}
in which
\begin{eqnarray}
U &=& -\frac{1}2 m_\omega^{*2}\omega^2 +\frac{m^2m_q^{*2}}{2g_s^2} +\frac{\lambda m_q^{*4}}{24g_s^4},
\nonumber \\
m_q^* &=& m_q -g_s u, \;\;\;\;\; m_q = g_s\sigma_0,
\label{3} \\
m_\omega^* &=& m_\omega -g_{sv} u, \;\;\;\;\; m_\omega = g_{sv}\sigma_0,
\nonumber \\
\mu^* &=&  \mu - g_v\omega_0.
\nonumber
\end{eqnarray}

Effective potential  derived from (\ref{2}) reads
\begin{eqnarray}
\Omega = U - \frac{6}{\pi^2} \int\!\! p^2 dp \left[ E_p + T\ln\left( 1+e^{-E_+/T}\right) \right.\nonumber\\
 \left.+ T\ln\left( 1+e^{-E_-/T}\right)\right]
\label{4}
\end{eqnarray}
where
\begin{eqnarray}
E_{\pm}(\vec{p}) = E_p \pm \mu^*, \;\;\;\;\;\;\; E_p = \sqrt{\vec{p}^2 +m_q^{*2}}.
\label{5}
\end{eqnarray}

The ground state of  the system is determined by  minimizing   (\ref{4}) 
\begin{eqnarray*}
\frac{\partial \Omega}{\partial u} = 0, \;\;\;\;\; 
\frac{\partial \Omega}{\partial \omega_0} = 0
\end{eqnarray*}
yielding
\begin{eqnarray*}
\frac{\partial U}{\partial u} = g_s \rho_s, & &\;\;\;\;\;\;\;
\rho_s = \frac{6}{\pi^2}\int\!\! p^2dp\ \frac{m_q^*}{E_p} (n_- +n_+ -1), \\
\frac{\partial U}{\partial \omega_0} = g_v \rho_v, & & \;\;\;\;\;\;\;
\rho_v = \frac{6}{\pi^2}\int\!\! p^2dp\ (n_- -n_+), \\ 
& &\;\;\;\;\;\;\; n_\pm = [1 + e^{E_\pm/T}]^{^{-1}},
\end{eqnarray*}
the first of which is usually called the gap equation. Combining the gap equation and   (\ref{3})  gives
\begin{eqnarray}
g_v^2 \frac{m_q^2}{m_\omega^2 m_q^{*3}}\ \rho_v^2 - \frac{m^2 m_q^*}{g_s^2} - \frac{\lambda m_q^{*3}}{6g_s^4}  = \rho_s.
\label{6}
\end{eqnarray}

Next let us establish the formulas for thermodynamic quantities, the first one among them is pressure $P$  to be determined through $\Omega$  taken at minimum
\[
P = -\Omega.
\]
The quark density is defined as
\begin{eqnarray*}
\rho = \frac{\partial P}{\partial \mu}
\end{eqnarray*}
giving
\begin{eqnarray}
\rho = \rho_v.
\label{7}
\end{eqnarray}

In term of quark density the expression for $P$ is written
\begin{eqnarray}
P &=& \frac{m^2 m_q^{*2}}{2g_s^2} + \frac{\lambda m_q^{*4}}{24g_s^4} - g_v^2\ \frac{m_q^2}{2 m_\omega^2 m_q^{*2}}\ \rho_v^2 \nonumber\\
& &\!\!\!+ \frac{6}{\pi^2} \!\!\int\!\!\! p^2 \!dp\! \left[ E_p \!+\! T\ln\left( \!1 \!+\! e^{-E_+/T}\!\right) \!+\! T\ln\left( \!1 \!+\! e^{-E_-/T}\!\right)\!\right]. \nonumber\\
\label{8}
\end{eqnarray}
The energy density $\epsilon$ is then obtained by a Legendre  of  $P$
\[
\epsilon = -P +Ts +\mu\rho
\]
which together with (\ref{8}) provides
\begin{eqnarray}
\epsilon &=& \frac{m^2 m_q^{*2}}{2g_s^2} + \frac{\lambda m_q^{*4}}{24g_s^4} + g_v^2\ \frac{m_q^2}{2 m_\omega^2 m_q^{*2}}\ \rho_v^2 \nonumber\\
& &+ \frac{6}{\pi^2}\int\!\! p^2dp\ E_p (n_- +n_+ -1).
\label{9}
\end{eqnarray}
Here
\begin{eqnarray}
s = \frac{6}{\pi^2}\int\!\! p^2dp\ [n_-\ln n_- +(1\!-\! n_-)\ln(1\!-\! n_-) \nonumber\\
+n_+\ln n_+ +(1\!-\! n_+)\ln(1\!-\! n_+)].
\label{10}
\end{eqnarray}

Eqs. (\ref{8}) and (\ref{9}) are the basic equations of state of  the quark system under    consideration.

At zero temperature Eqs. (\ref{6}), (\ref{7}), (\ref{8}) and (\ref{9}) reduce to 
\begin{eqnarray}
& & g_v^2 \frac{m_q^2}{m_\omega^2 m_q^{*3}}\ \rho^2 \!-\! \frac{m^2 m_q^*}{g_s^2} \!-\! \frac{\lambda m_q^{*3}}{6g_s^4}  = \frac{6}{\pi^2}\!\int_0^{p_{_F}}\!\!\! p^2 \!dp\ \frac{m_q^*}{E_p},
\label{11}\\
& & \rho = \frac{6 p_{_F}^3}{3\pi^2},
\label{12}\\
& & P = \frac{m^2 m_q^{*2}}{2g_s^2} \!+\! \frac{\lambda m_q^{*4}}{24g_s^4} \!-\! g_v^2 \frac{m_q^2}{2 m_\omega^2 m_q^{*2}} \rho^2 \!+\! \frac{6}{\pi^2}\! \int_0^{p_{_F}}\!\!\! p^2\! dp E_p,
\label{13}\\
& & \epsilon = \frac{m^2 m_q^{*2}}{2g_s^2} \!+\! \frac{\lambda m_q^{*4}}{24g_s^4} \!+\! g_v^2 \frac{m_q^2}{2 m_\omega^2 m_q^{*2}} \rho^2 \!+\! \frac{6}{\pi^2}\!\int_0^{p_{F}}\!\!\! p^2\! dp E_p.
\label{14}
\end{eqnarray}

With the aid of the analytical expressions derived in the foregoing Section let us investigate successively the following phase transitions.

\section{Chiral symmetry non-restoration}

For the numerical calculations let us use the inputs such as $g_s = 3.3$ fm$^2$, $g_v = 0.99$ fm$^2$, $m_q = 5$ MeV, $m_\sigma = 600$ MeV, $m_\omega = 783$ MeV.

Let us first numerically compute the $T$ dependence of  the order parameter $m_q^*$   at  $\mu = 0$ (Fig.~\ref{f1}) and $\mu =$ 100, 160, 200, 220 MeV (Fig.~\ref{f2}). The numerical computation implemented for  Eqs. (\ref{6}) and (\ref{11}) produces Figs.~\ref{f1} and \ref{f2}  with the main properties: Fig.~\ref{f1}  shows that at $\mu = 0$ the order parameter decreases to zero as $T$ increases infinitely. Thus, the chiral symmetry gets  restored and the behavior of  the graph means that the transition is first order. Meanwhile, at non-vanishing $\mu$  Fig.~\ref{f2}  yields the expected scenario of chiral non-restoration. 

\begin{figure}[t] 
\begin{minipage}{\columnwidth}
\centering
\includegraphics[width=\columnwidth,height=6cm]{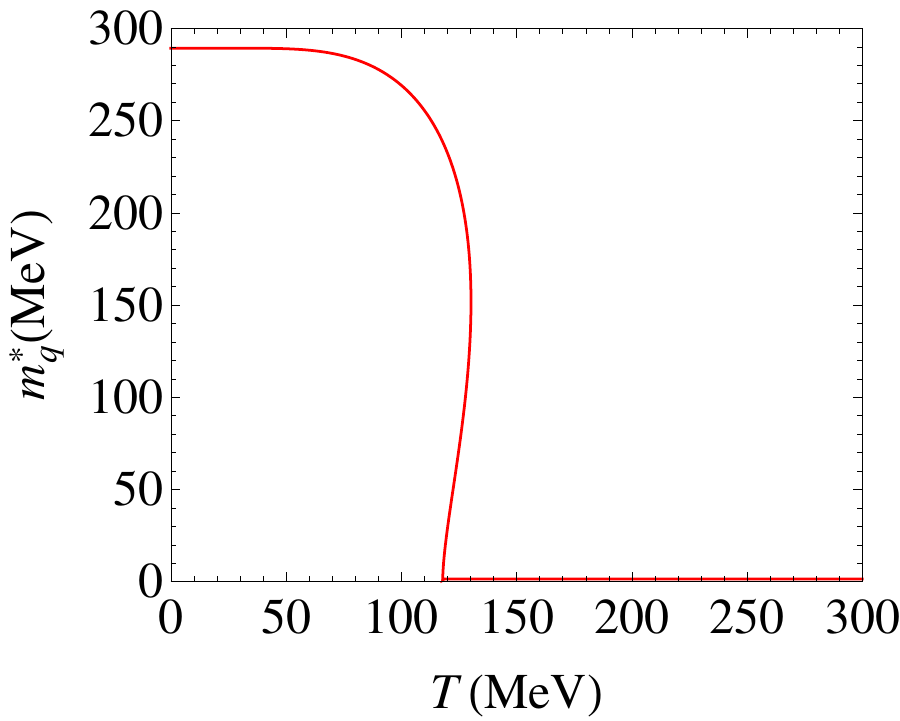}
\end{minipage}
\caption{The $T$ dependence of  order parameter at  $\mu = 0$.} 
 \label{f1}
\end{figure}
\begin{figure}[h] 
\begin{minipage}{\columnwidth}
\centering
\includegraphics[width=\columnwidth,height=6cm]{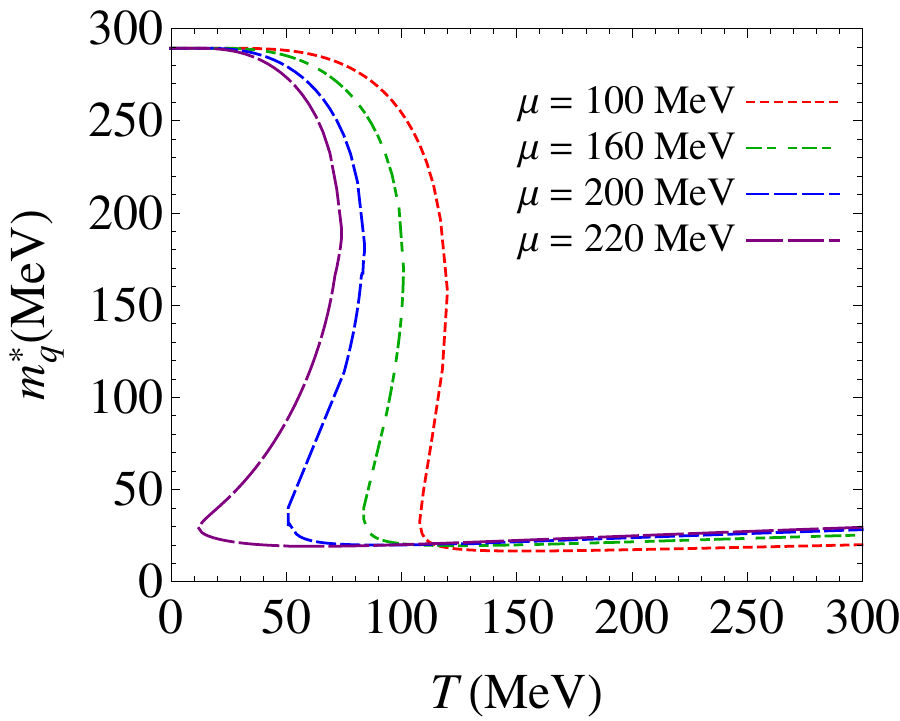} 
\end{minipage}
\caption{The $T$ dependence of order parameter at $\mu =$ 100 MeV (dotted line), 160 MeV (dashed-dotted line), 200 MeV (dashed line) and 220 MeV (long dashed line). } 
 \label{f2}
\end{figure}

Next the $\mu$ dependence of order parameter $m_q^*$  at  $T = 0$ (solid line), 115 MeV (dashed line)  is  plotted in Fig.~\ref{f3}. The behavior of  $m_q^*$   typically describes the first - order phase transition in the region restricted by two dotted lines where it is a multi-valued function of  $\mu$. The isothermal spinodal points which delimit  the region of  equilibrium first-order phase transition  are denoted by A, A' and B, B'. It is easily seen that the chiral symmetry is  not restored at  high  baryon density, too. As we will see later,  the feature  that $m_q^*$  is a multi-valued function of $\mu$ in a large range of  $T$  gives rise to  the specific behaviors of energy density and pressure as functions of  $\mu$  in a large range of  $T$.

\begin{figure}[h] 
\begin{minipage}{\columnwidth}
\centering
\includegraphics[width=\columnwidth,height=6cm]{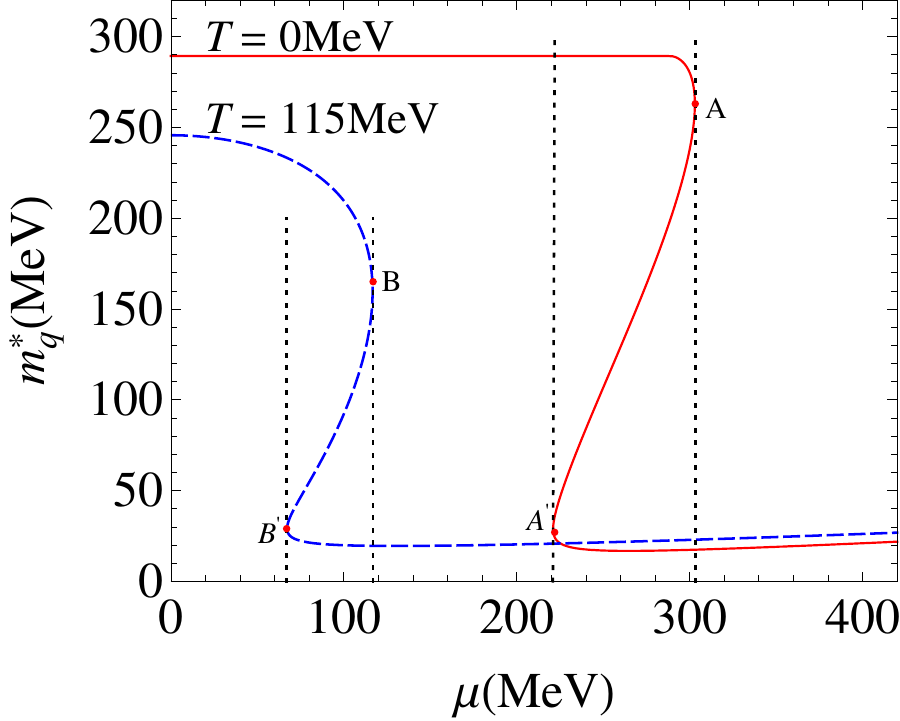} 
\end{minipage}
\caption{The $\mu$ dependence of  order parameter at $T = 0$ (solid line), 115 MeV (dashed line). The region of the first-order phase transition is restricted by two dotted lines. A, A' and B , B' are isothermal spinodal points.}
\label{f3}
\end{figure}

\section{Lifshitz phase transition}

It is commonly accepted that  there exist so far two schemes of classifications relevant  to physical systems:
\begin{enumerate}
\item[a)] The conventional classification by symmetry which reflects the phenomena of spontaneously broken symmetry when the energy is reduced.
\item[b)] Another scheme of classification deals with classifying vacuum states by momentum-space topology \cite{r10, r11}. It reflects the opposite tendency, the symmetry gradually emerges at low energy. The universality classes of quantum vacuum states are determined by the momentum-space topology, which is primary while the vacuum symmetry is the emergent phenomenon in the low energy corner.
\end{enumerate}

For all fermionic systems in three-dimensional (3D) space and invariant under translations of coordinates there are four basics universality classes of vacua provided by topology in momentum space \cite{r10, r11}:
\begin{enumerate}
\item[-] Vacua with fully-gapped fermionic energy excitations, such as semiconductors, superconductors and Dirac particles.
\item[-] Vacua with fermionic energy excitations characterized by Fermi points in 3D momentum space determined by  $E(\vec{p})=0$.
\item[-] Vacua with fermionic energy excitations determined by Fermi surfaces in 3D momentum space, and
\item[-] Vacua with fermionic energy excitations characterized by lines in the 3D momentum space.
\end{enumerate}

The phase transitions which follow from this classification scheme are quantum phase transition occurring at $T = 0$. Among these QPT's  the Lifshitz phase transition (LPT) is marked as a QPT from the fully-gapped state to the state with Fermi surface \cite{r12}, which should be not confused with the Lifshitz points \cite{r13}. In recent years LPT becomes one of the hot topics in condensed matter physics \cite{r14, r15, r16, r17}. The LPT in nuclear matter  has been studied in Refs.~\cite{r18, r19}. 

Now we present the LPT in the effective model (\ref{1}) of  QCD characterized by the chiral non-restoration at high temperature and/or baryon chemical potential. We first concern the case $T = 0$, then extend to the finite temperature case.

In the vacuum state the quark energy excitation possesses the form (\ref{5}),
\[
E_{_\pm}(\vec{p}) = \sqrt{\vec{p}^2 + m_q^{*2}} \pm \mu^*
\]
Then the locus of Fermi points are determined by
\begin{eqnarray*}
E_{_+}(\vec{p}) = 0 & &\mbox{for } \mu^*>0 \\
E_{_-}(\vec{p}) = 0 & &\mbox{for } \mu^*<0 
\end{eqnarray*}

Let us first consider the case $\mu^* >0$  where the locus of Fermi points in the  ($p_z, \mu$)- plane is given by the equation
\begin{eqnarray}
p_z = \pm \sqrt{\mu^{*2} - m_q^{*2}}, \;\;\;\;\;\;\; p_x = p_y = 0.
\label{15}
\end{eqnarray}

\begin{figure}[h] 
\begin{minipage}{\columnwidth}
\centering
\includegraphics[width=\columnwidth,height=6cm]{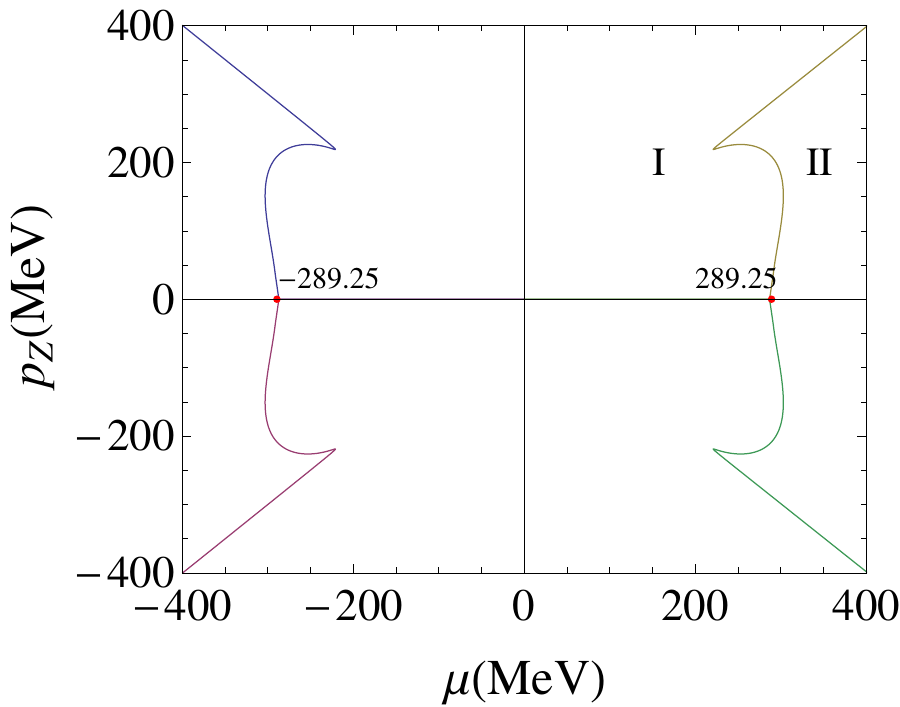} 
\end{minipage}
\caption{The locus of Fermi points in the  ($k_z, \mu$)- plane in the case $\mu^* >0$.}
\label{f4}
\end{figure}

In the Fig.~\ref{f4} is shown this locus of  Fermi point in the ($p_z, \mu$)-plane which is the boundary separating the domain I  where $\mu^{*2} - m_q^{*2} <0$ from the domain II with   $\mu^{*2} - m_q^{*2} >0$. In 4-dimensional space ($p_x, p_y, p_z, \mu$)    the locus of  Fermi points forms the Fermi sphere in 3-dimensional momentum space with center in the $\mu$  axis and radius equal to $p_{_F} = \sqrt{\mu^{*2} - m_q^{*2}}$. Close to this sphere the spectrum of  quasi-fermion takes the form 
\begin{eqnarray}
E_{_-} \approx \frac{\vec{p}^2 - p_{_F}^2}{2\mu^*} = \frac{\vec{p}^2}{2\mu^*} - \epsilon_{_F}.
\label{16}
\end{eqnarray}
Here $\epsilon_{_F}= p_{_F}^2/2\mu^*$ is the Fermi energy . Thus , the domain II corresponds to the states with Fermi sphere . In the domain I  the spectrum of quasi-fermions possesses non-vanishing energy gap $E_{_-}(\vec{p})= m_q^{*2} - \mu^{*2} \neq 0$. It is the domain of fully-gapped states. The point $\mu = \mu_0$ = 289.25 MeV corresponding to  $\mu^{*2} - m_q^{*2} = 0$ is exactly the critical point  of the LPT  from fully-gapped states to states with Fermi sphere.

It is known that for a general system, relativistic or non-relativistic, the stability of the $j$-th Fermi point is guaranteed by the topological invariant, which is called the winding number $N_j$   defined as
\begin{eqnarray}
N_j = \frac{1}{24\pi^2}\ \epsilon_{_{\mu\nu\rho\sigma}}\mbox{ tr} \!\oint_{\Sigma_j} \!\!dS^\sigma\frac{\partial}{\partial p_\mu} G^{-1}\!G \frac{\partial}{\partial p_\nu} G^{-1}\!G \frac{\partial}{\partial p_\rho} G.
\label{17}
\end{eqnarray}
Here $G$ is the quark propagator, $\Sigma_j$  is the 2D surface around the $j$-th isolated Fermi point $E_{_-}(\vec{p})=0$, $j = 1$, 2, and the trace is taken over all relevant indices. In our case
\[
G^{-1} = i\omega -H(\vec{p}), \;\;\;\;\;\;\; H(\vec{p}) = \vec{\alpha}\vec{p} +\mu^* -\beta m_q^*
\]
which, combining  with (\ref{15}) and (\ref{17}), yields $N_1 = +1$, $N_2 = -1$. At  $\mu = \mu_0$  these two points merge and form one topological trivial Fermi point $\vec{p}_0=0$  with winding number $N_0 =N_1 +N_2 =0$. This intermediate state is marginal and cannot protect the vacuum against decay into two topologically stable vacua.

It is evident that the  stability of two Fermi points entails the stability of  the Fermi sphere. However, the Fermi sphere itself is also a topologically stable  object protected by another  winding number. Indeed, in the domain II we have the inverse Green function (in imaginary frequency)
\[
G^{-1} = i\omega - \frac{\vec{p}^2}{2\mu^*} + \varepsilon_{_F}
\]
taking into account (\ref{16}).

The stability of the Fermi sphere is then guaranteed by the non-vanishing winding number
\[
N_{_{FS}} = \mbox{ tr} \oint_{\mbox{\tiny around FS}} \frac{dp_\mu}{2\pi i} G \frac{\partial}{\partial p_\mu} G^{-1}= 1.
\]
As for the energy excitation $E_+(\vec{p})$  the locus of Fermi points corresponds to $\mu<0$. As is shown in Fig.~\ref{f4} the loci of Fermi points associating with $E_+(\vec{p})$ and $E_-(\vec{p})$  are symmetric through the $p_z$  axis. Hence, for the sake of simplicity we focus on the case $\mu > 0$ only.

At finite temperature $m_q^*$ and $\mu^*$   turn out to be functions of both  $T$  and $\mu$. As a matter of fact, due to (\ref{15}) the states with Fermi sphere are determined by
\[
\theta(T,\mu) = \mu^{*2}(T,\mu) - m_q^{*2}(T,\mu) > 0
\]
with two Fermi points 
\[
p_z(T,\mu) = \pm \sqrt{\mu^{*2}(T,\mu) - m_q^{*2}(T,\mu)}, \;\;\;\;\;\;\; p_x = p_y = 0.
\]
The fully-gapped states are characterized by  $\theta(T,\mu)<0$  and, consequently, the critical states are given by 
\begin{eqnarray}
\theta(T,\mu) = \mu^{*2}(T,\mu) - m_q^{*2}(T,\mu) = 0
\label{18}
\end{eqnarray}
separating fully-gapped states from states with Fermi sphere. This means that  $\theta$ plays the role of  order parameter of  LPT  and Eq.~(\ref{18}) describes  the phase diagram of  LPT in the $(T, \mu)$-plane  plotted in  Fig.~\ref{f5}.
\begin{figure}[h] 
\begin{minipage}{\columnwidth}
\centering
\includegraphics[width=\columnwidth,height=6cm]{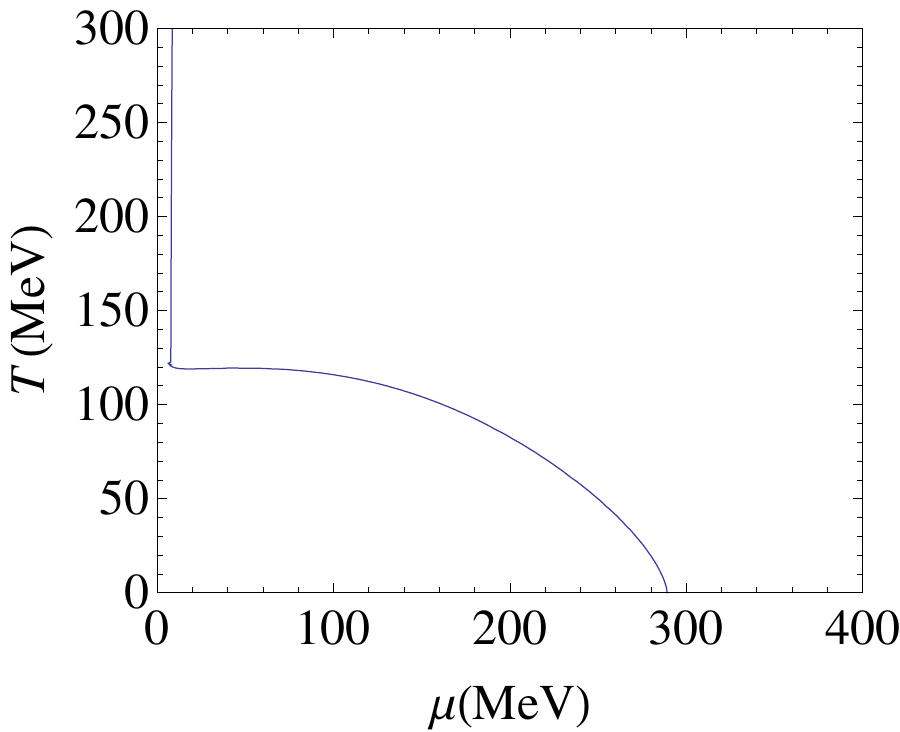} 
\end{minipage}
\caption{Phase diagram of  LPT. The phase transition is  first order everywhere.}
\label{f5}
\end{figure}
\begin{figure}[h] 
\begin{minipage}{\columnwidth}
\centering
\includegraphics[width=\columnwidth,height=6cm]{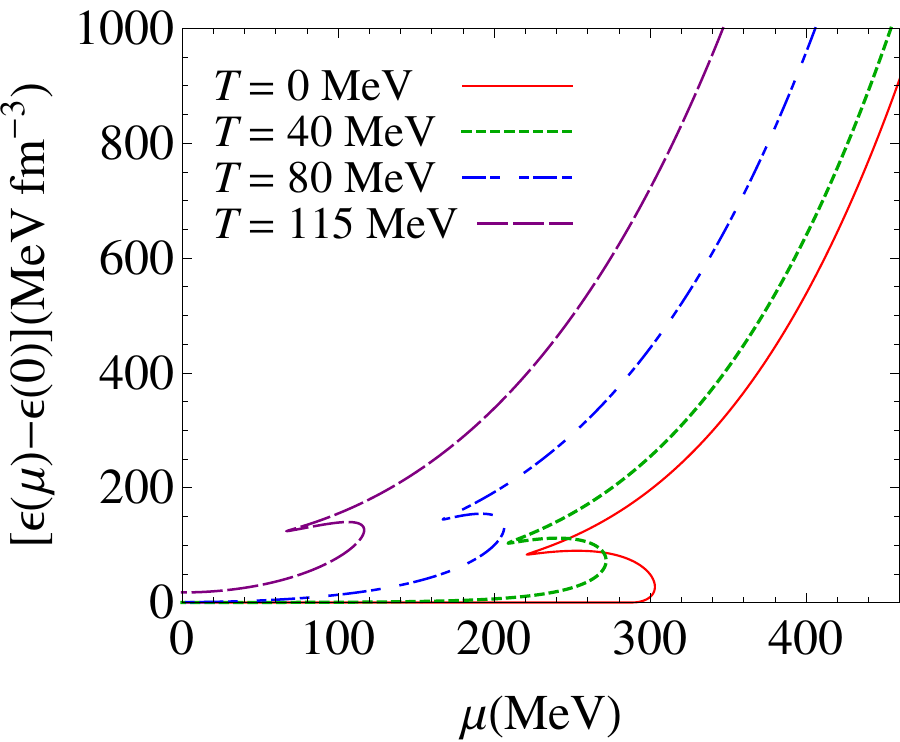} 
\end{minipage}
\caption{The evolution of $\epsilon$ versus $\mu$ at  $T = 0$ (solid line), 40 MeV (dotted line), 80 MeV (dashed-dotted line) 115 MeV (dashed line).}
\label{f6}
\end{figure}

In order to determine the order of LPT let us examine the $\mu$ dependence of  energy density, Eqs.~(\ref{9}) and (\ref{14}),  at several values of $T =$ 0, 40, 80, 115 MeV. Its graphs  depicted in Fig.~\ref{f6}  are featured by the existence of cusps and their behaviors which are typical for first-order phase transition: during phase transition the system emits an amount of latent heat.  For example, at $T = 0$ the QCP of  LPT  locates at   $\mu = 289.25$ MeV, from  Fig.~\ref{f6} the emitted latent heat  is derived
\begin{eqnarray*}
\Delta \epsilon &=& \epsilon (\mu \!=\! 289.25 \mbox{ MeV} + 0, T \!=\! 0) \\
& & - \epsilon (\mu \!=\! 289.25 \mbox{ MeV} - 0, T \!=\! 0) \\
&=& 138.50 \mbox{ MeV}^{-3}\mbox{fm}.
\end{eqnarray*}

Hence, we conclude that  the LPT is first order.

\begin{figure}[h] 
\begin{minipage}{\columnwidth}
\centering
\includegraphics[width=\columnwidth,height=6cm]{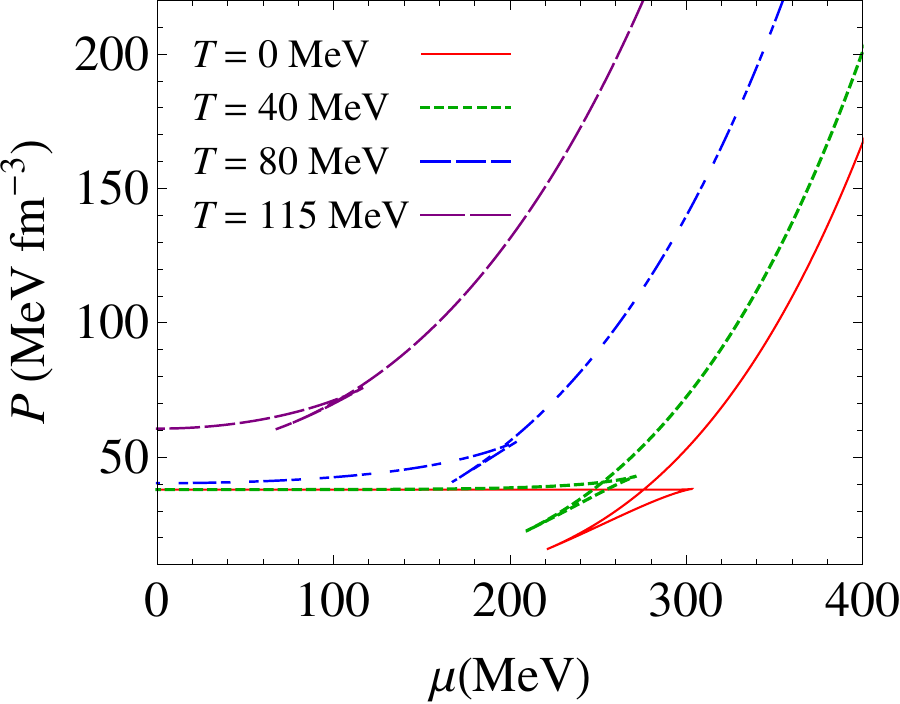} 
\end{minipage}
\caption{The evolution of $P$ versus $\mu$ at  $T = 0$ (solid line), 40 MeV (dotted line), 80 MeV (dashed-dotted line) 115 MeV (dashed line).}
\label{f7}
\end{figure}

In addition, according to Abrikosov \cite{r20} the order of LPT is also derived  by considering the  deviation of energy density $\epsilon$ and pressure $P$ from their critical values. The evolution of  $P$, Eqs.~(\ref{8}) and (\ref{13}), versus $\mu$ at several temperature steps plotted in Fig.~\ref{f7}. Our calculation provides   
\[
\Delta\epsilon \approx 1.26 |\mu -\mu_c|^{0.99} \approx 5 \Delta P, \;\;\;\;\;\;\; \mu_c = 289.25 \mbox{ MeV},
\]
which means that  LPT  is  first order.

In order to get deeper insight into the issue  let us observe the evolution of  $\partial\epsilon/\partial\mu$  at  several values of  $T$  depicted in Fig.~\ref{f8}. The analytical formula for  this quantity is given in Appendix. Combining  Figs.~\ref{f6} and \ref{f8} indicates that for given $T$   the graph of  $\epsilon$  possesses two bifurcation points: the first one corresponds to the divergence of  $\partial\epsilon/\partial\mu$  and the second one is exactly the cusp where $\partial\epsilon/\partial\mu$   is not well definite, its value depends on the way we approach this cusp. As we will see later, the  segment connecting  these two bifurcation points belongs to the mechanically unstable region.

The existence of cusps also appears in the  evolution of the pressure $P$ shown in Fig.~\ref{f7} which tells  that  at every value of $T$  there exhibit two cusps, one of them corresponds to the cusp of $\epsilon$  and another cusp corresponds to the second bifurcation point of $\epsilon$.
\begin{figure}[h] 
\begin{minipage}{\columnwidth}
\centering
\includegraphics[width=\columnwidth,height=6cm]{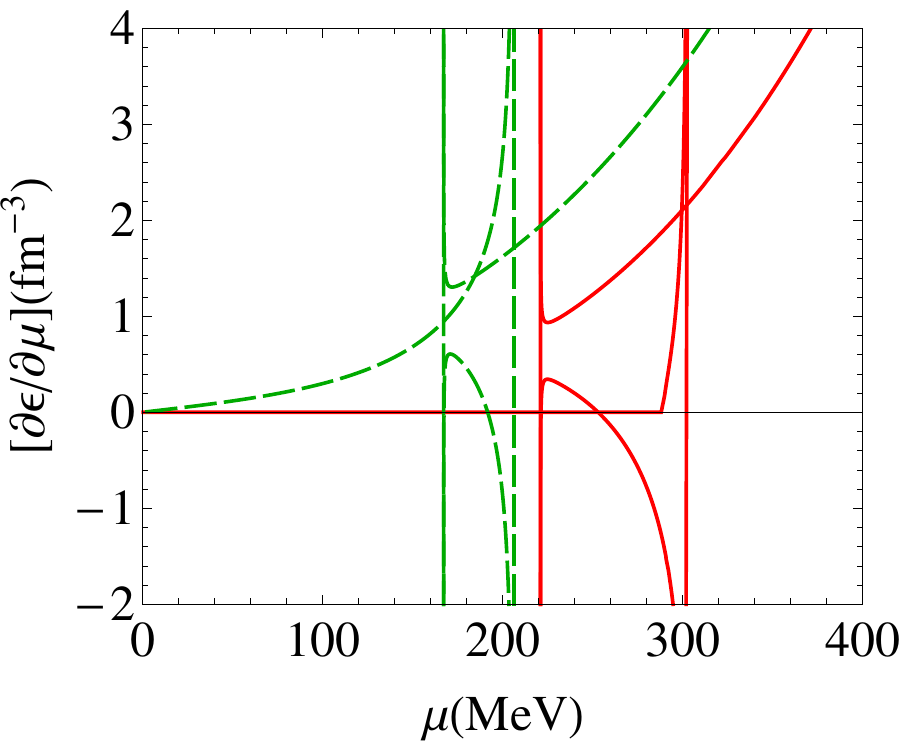} 
\end{minipage}
\caption{The evolution of  $\partial\epsilon/\partial\mu$   versus $\mu$ at $T = 0$ (solid line) and 80 MeV (dashed line).}
\label{f8}
\end{figure}

Another feature of LPT is cleared up by  investigating the evolution of  quark density  against $\mu$ at $T = 0$, 40, 80, 115 MeV. Eqs.~(\ref{7}) and (\ref{12}) together with the gap equation provide Fig.~\ref{f9} which implies that at fixed $T$ as $\mu$ changes from $\theta(T,\mu) <0$ to  $\theta(T,\mu) >0$ the system undergoes a transition from  fully-gapped state with low density  to  a state with Fermi sphere at higher density. At $\theta(T,\mu) =0$   the system has a jump from low to high densities. If  we identify these two states to the gas and liquid states accordingly,  then  LPT  describes  a quantum liquid-gas phase transition  from fully-gap state to state with Fermi sphere. At $T = 0$ the quark system resides in the state with Fermi sphere where the formula (\ref{12}) is applied.
\begin{figure}[h] 
\begin{minipage}{\columnwidth}
\centering
\includegraphics[width=\columnwidth,height=6cm]{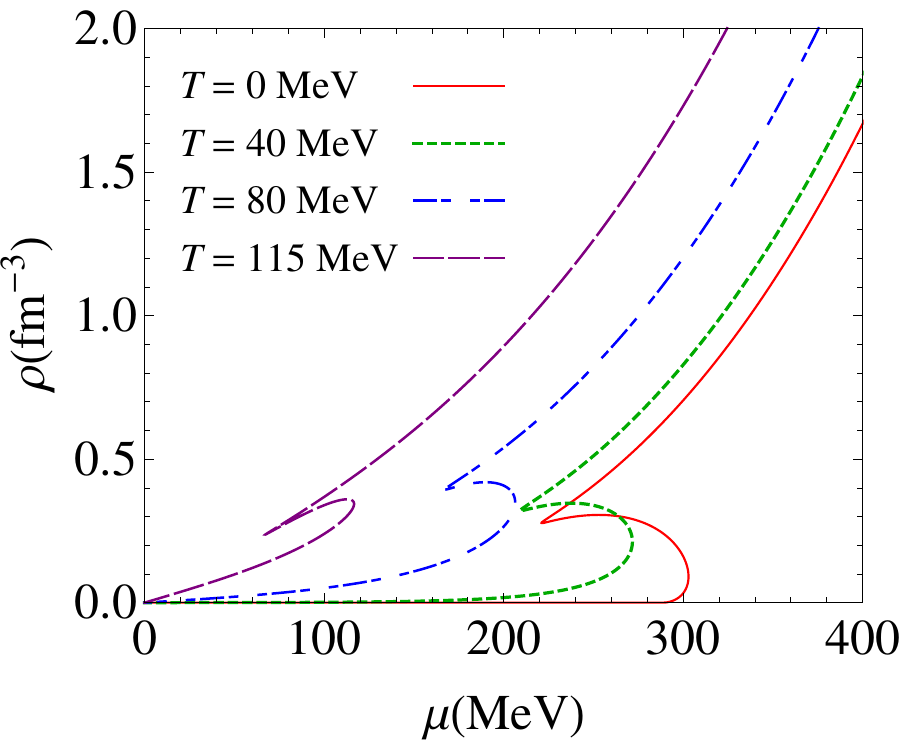} 
\end{minipage}
\caption{The evolution of  quark density    versus $\mu$ at $T = 0$ (solid line), 40 MeV (dotted line), 80 MeV (dashed-dotted line) 115 MeV (dashed line).}
\label{f9}
\end{figure}

To proceed to the stability of states in LPT  let us  introduce the density    susceptibility  defined as
\[
\chi = \frac{\partial \rho}{\partial\mu} =-\frac{\partial^2\Omega}{\partial\mu^2}
\]
whose analytical formula is presented in Appendix. It is commonly accepted that the density susceptibility is related to density fluctuation which allows us to observe signals of phase transition in experiments. We show in Fig.~\ref{f10} the graphs describing the evolution of $\chi$ versus $\mu$ at  several values of  $T$. Let us analyze some characters exhibiting in this figure. 

For  $T = 0$ we see that $\chi$  diverges at  $\mu = 302.4$  MeV and has three branches in the interval 250.10 MeV $\leq\mu\leq$ 302.40 MeV: two positive branches correspond to stable states which associate with minima of effective potential and the negative one corresponds to mechanically unstable states where  effective potential gets maximum, therefore,  all points belonging to this branch describe the spinodal line. Two coexisting states in the LPT, occurring at   $\mu_0 =$ 289.25 MeV,  belong to the topologically stable states: the one with low density is  the fully-gapped state which is protected against any small perturbation \cite{r11} and the other with higher density belongs to the domain of state with Fermi sphere protected by the winding number $N_{_{FS}}$. At finite $T$  Fig.~\ref{f10} exhibits  the same physical picture. 
\begin{figure}[h] 
\begin{minipage}{\columnwidth}
\centering
\includegraphics[width=\columnwidth,height=6cm]{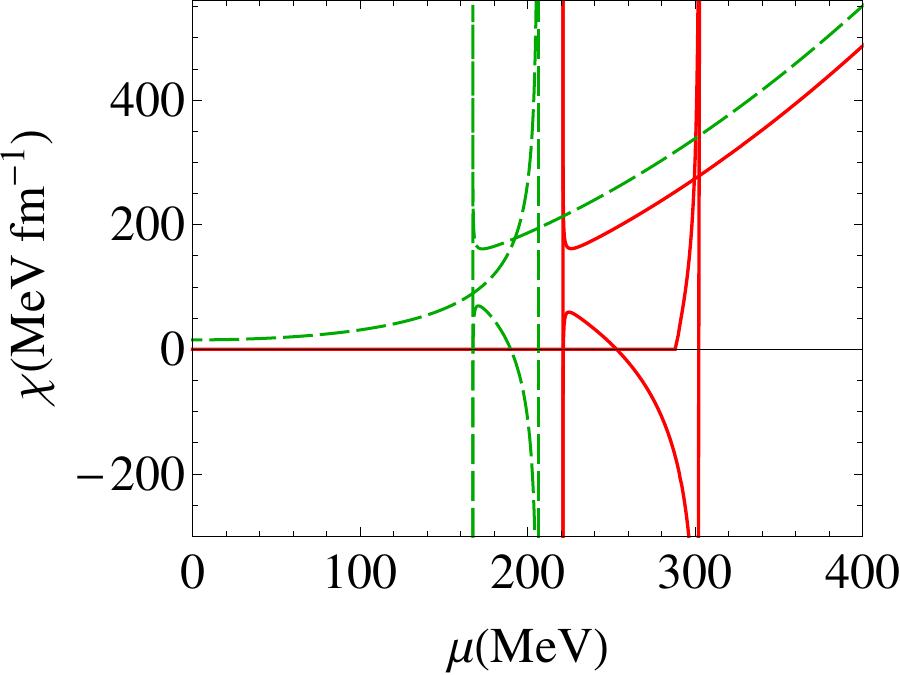} 
\end{minipage}
\caption{The evolution of  the density susceptibility  versus $\mu$ at $T = 0$ (solid line) and 80 MeV (dashed line).}
\label{f10}
\end{figure}

It is easy to recognize a common property of  Figs.~\ref{f6} and \ref{f8} which express the fact that for every value of $T$  two bifurcation points of  $\epsilon$ and $\rho$ emerge at the same value of $\mu$. In view of the foregoing result, the segment to connect these two bifurcation points of $\epsilon$  belongs to the mechanically unstable region. Analogously, at given $T$ the segment connecting two cusps  of  pressure plotted in Fig.~\ref{f7}  is exactly the domain of  mechanically unstable states.  Fig.~\ref{f9}  indicates  that each  unstable state is the  intermediate one separating  a state with higher density from state with lower density.

\section{Conclusion and outlook}

In the preceding Section we presented the topological Lifshitz phase transition in an effective model of QCD where the chiral symmetry broken at $T = 0$  never gets restored at high $T$ and/or  $\mu$. The main results we found are in order
\begin{enumerate}
\item[a)] We indicated that under the LPT the quark system undergoes a transition from liquid to gas states which correspond respectively to fully-gapped state and  state with Fermi sphere. In this connection, the gas state is stable under any small perturbation \cite{r10, r11} and the liquid state is protected by non-vanishing values of winding numbers.
\item[b)] We obtained the Lifshitz phase diagram in the ($T, \mu$)-plane. The transition is first order everywhere.
\end{enumerate}

It is interesting to mention that  the calculations performed for the Nambu--Jona-Lasinio model \cite{r21, r22, r23, r24} predict that the chiral restoration at  $\mu \neq 0$ is of rather strong first order at  $T < 50$ MeV, provided that the vector coupling between quarks  is absent. As a matter of fact, the strength and even the existence of the  first-order restoration transition  strongly depend on the strength of the vector coupling. Hence, in this case we also have another effective model with chiral non-restoration and  what we found  above is valid for  this class of  models, whose quark  energy  excitation takes the form
\[
E = \sqrt{\vec{p}^2 + M^2(T,\mu)} - F(T,\mu),
\]
$M$ and $F$ are functions of control parameter denoted by $\mu$ and other parameters involving temperature $T$. Then the order parameter of LPT is defined as 
\[
\theta = F^2 - M^2.
\]
Depending on $\mu$ the order parameter can get various values from $\theta>0$ to $\theta>0$. It results that corresponding to negative $\theta$ is the fully-gapped state and the positive $\theta$ corresponds to state with Fermi surface which is protected by momentum-space topology.

\begin{acknowledgements}
T.H.P. is supported by  the MOST  through the Vietnam - France LIA    collaboration. P.T.T.H  and N.T.A. are supported by  NAFOSTED under grant No. 103.01-211.05.
\end{acknowledgements}

\appendix

\section{The analytical expression for  $\partial\epsilon/\partial\mu$ and $\chi$}

\begin{eqnarray*}
& & \frac{\partial \epsilon}{\partial\mu} = \frac{G B + H D}{B C + D E}, \\
& & \chi = \frac{A B + F D}{B C + D E},
\end{eqnarray*}
where
\begin{eqnarray*}
& & A = \frac{2m_q^*}{\pi^2 T}\ I_5, \\
& & B = \frac{2g_v^2 m_q^2}{\pi^4 T m_\omega^2 m_q^{*2}}\ I_1 I_3 - \frac{m_q^{*2}}{2\pi^2T}\ I_5, \\
& & C = \frac{4g_v^2 m_q^2}{\pi^2 m_\omega^2 m_q^{*3}}\ I_1 +  \frac{2g_v^2m_q^2}{\pi^2 T m_\omega^2 m_q^{*}}\ I_5,\\
& & D = -\frac{m_\sigma^2 m_q^{*3}}{4g_s^2 m_q^2} + \frac{m_\sigma^2 m_q^* (m_q^2 - m_q^{*2})}{4g_s^2 m_q^2} + \frac{2g_v^2 m_q^2}{\pi^4 m_\omega^2 m_q^{*3}}\ I_1 \\
& &\hspace{.8cm} -\frac{m_q^*}{2\pi^2}\ I_2 + \frac{m_q^*}{2\pi^2}\ I_9 - \frac{2g_v^2 m_q^2}{\pi^4 T m_\omega^2 m_q^{*}}\ I_1 I_5,
\\
& & E= 1 + \frac{2g_v^2 m_q^2}{\pi^2 T m_\omega^2 m_q^{*2}}\ I_3,
\\
& & F = \frac{2}{\pi^2 T}\ I_3,
\\
& & G = -\frac{m_\sigma^2 m_q^* (m_q^2 - m_q^{*2})}{2g_s^2 m_q^2} + \frac{4g_v^2 m_q^2}{\pi^4 m_\omega^2 m_q^{*3}}\ I_1 + \frac{2m_q^*}{\pi^2}\ I_2  \\
& &\hspace{.8cm} -\frac{2m_q^*}{\pi^2 T}\ I_3  - \frac{4g_v^2 m_q^2}{\pi^4 T m_\omega^2 m_q^{*}}\ I_1 I_5,\\
& & H = - \frac{4g_v^2 m_q^2}{\pi^4 T m_\omega^2 m_q^{*2}}\ I_1 I_3 - \frac{2}{\pi^2T}\ I_4, 
\\
& & P = - \frac{2m_q^{*}}{\pi^2T}\ I_8, 
\\
& & Q = -\frac{2g_v^2 m_q^2}{\pi^4 T m_\omega^2 m_q^{*2}}\ I_1 I_6 + \frac{m_q^{*2}}{2\pi^2T^2}\ I_8, 
\\
& & K = \frac{2g_v^2 m_q^2}{\pi^2 T^2 m_\omega^2 m_q^{*2}}\ I_6, 
\\
& & N = - \frac{2}{\pi^2 T}\ I_6,
\\
& & M = \frac{2}{\pi^2 T^2}\ I_7,\\
\end{eqnarray*}
in which the integrals as follows,
\begin{eqnarray*}
& & I_1 = \int dk\ k^2(n_+ - n_-),\\
& & I_2 = \int dk\ \frac{k^2}{E_k}(n_+ + n_-),\\
& & I_3 = \int dk\ k^2(n_+ + n_- -n_+^2 -n_-^2),\\
& & I_4 = \int dk\ k^2 E_k (n_+ - n_- -n_+^2 +n_-^2),\\
& & I_5 = \int dk\ \frac{k^2}{E_k} (n_+ - n_- -n_+^2 +n_-^2),\\
& & I_6 = \int dk\ k^2 [E_+ (n_+ -n_+^2) - E_-(n_- -n_-^2)],\\
& & I_7 = \int dk\ k^2 [E_+^2 (n_+ -n_+^2) + E_-^2 (n_- -n_-^2)],\\
& & I_8 = \int dk\ \frac{k^2}{E_k} [E_+ (n_+ -n_+^2) + E_-(n_- -n_-^2)],\\
& & I_9 = \int dk\ \frac{k^2}{E_k^2} \bigg[\frac{m_q^{*2}}T (n_+ +n_- -n_+^2 -n_-^2)  - \frac{k^2}{E_k} (n_+ +n_-)\bigg].
\end{eqnarray*}

\end{document}